\documentclass[12pt]{article}
\usepackage{amsmath}
\usepackage{amssymb,amsfonts}
\usepackage{hyperref}              
\providecommand{\href}[2]{#2}
\renewcommand{\baselinestretch}{1.2}
\def\sp{\;\;\;,\;\;\;}

\def\e{\epsilon}

\def\m\mu
\def\n{\nu}

\newcommand{\eis}[3]{~{\cal E}^{#1}_{\irrep{#2};#3}}

\newcommand{\so}[2]{~[SO(#1)\times SO(#2)] \backslash SO(#1,#2,\Real)}

\newcommand{\ar}[2]{\left[ {}^{#1}_{#2} \right]}

\newcommand{\irrep}[1]{{\bf #1}}
\newcommand{\exc}{\ensuremath{E_{d+1(d+1)}}}

\newcommand{\Tr}{{\rm Tr}}

\newcommand{\D}{{\cal D}}

\newcommand{\F}{{\cal F}}

\newcommand{\Zint}{\mathbb{Z}}
\newcommand{\Real}{\mathbb{R}}

\def\n{\nu}

\newcommand{\iia}{{\rm IIA}}

\newcommand{\lh}{l_{\rm H}}
\newcommand{\rh}{R_{\rm H}}

\newcommand{\lii}{l_{\rm II}}

\newcommand{\lp}{l_{\rm P}}
\newcommand{\gh}{g_{\rm H}}
\newcommand{\gii}{g_{\rm II}}

\newcommand{\giis}{g_{\rm 6IIA}}
\newcommand{\giit}{g_{\rm 3IIA}}
\newcommand{\ghs}{g_{\rm 6H}}

\newcommand{\ght}{g_{\rm 3H}}
\newcommand{\ghf}{g_{\rm 4H}}

\renewcommand{\sp}{\ ,\qquad}
\makeatletter 
\renewcommand{\@makefnmark}{\mbox{$^{\ddagger\@thefnmark}$}}
\renewcommand{\subsection}{\@startsection
  {subsection}{2}{0pt%-1em
}{-\baselineskip}{0.5\baselineskip}
  {\normalfont\normalsize\itshape}}
\renewcommand{\section}{\@startsection
  {section}{2}{0pt%-1em
}{-\baselineskip}{0.5\baselineskip}
  {\bf\large}}

\makeatother
\numberwithin{equation}{section}
\numberwithin{table}{section}

\newcommand{\publititle}[8]
{ %\clearemptydoublepage
  \vspace*{-3cm}
  \begin{flushright} #1 \\ {\tt #2} \end{flushright}
  \vfill
  \begin{center}{\Large%\scshape
    \bfseries #3}\end{center}
  \vskip 8mm
  \begin{center}{\large #4}\end{center}
  \begin{center}{\normalsize\sl #5}\end{center}
  \vskip 8mm
  \nopagebreak
  \noindent #6
  \vfill
  \begin{flushleft} #7
  \end{flushleft}
  \hrule width 6.7cm \vskip.1mm
  {\small #8}
  \thispagestyle{empty}
  \clearpage
}                           

\begin{document}

\publititle{SPIN-01/01 \\ ITP-UU-01/01 \\
 HUTP-00/A053 \\ LPTHE-01-51}{hep-th/0101122}  
{Exact Thresholds and Instanton Effects \\ in String Theory ${}^{\star}$} 
{N.A. Obers$^{\,a\dagger}$ and B. Pioline$^{\,b\ddagger}$} 
{
$^a$ Spinoza Institute, Utrecht University, 3584 CE Utrecht, The Netherlands
\\ {\it and} \\ Institute for Theoretical Physics, Utrecht University,  
3508 TA Utrecht, The Netherlands 
\\[2mm]
$^b$
Jefferson Physical Laboratory, Harvard University, Cambridge MA
02138, USA
\\ {\it and} \\  
LPTHE, Universit{\'e} Pierre et Marie Curie,
Paris VI \\   Universit{\'e} Denis Diderot, Paris VII, Bo\^{\i}te
126, Tour 16, 1$^{\it er}$ {\'e}tage, 4 place Jussieu, F-75252
Paris Cedex 05, France
} 
{ 
In this lecture we summarize some recent work on the understanding
of instanton effects in string theories with 16 supersymmetries.
In particular, we consider $F^4$ couplings using the duality
between the heterotic string on $T^4$ and type IIA on $K_3$ at an
orbifold point, as well as higher and lower dimensional versions
of this string-string duality. At the perturbative level a
non-trivial test of the duality, requiring several miraculous
identities, is presented by matching a purely one-loop heterotic
amplitude to a purely tree-level type II result. A wide variety of
non-perturbative effects is shown to occur in this setting,
including D-brane instantons for type IIA on $K_3 \times S^1$ and
NS5-brane instantons for type IIA on $K_3 \times T^2$. Moreover,
the analysis of the three-dimensional case, which possesses a
non-perturbative $SO(8,24,\Zint)$ U-duality, reveals the presence of
Kaluza-Klein 5-brane instanton effects, both on the heterotic and
the type II side.
} 
{January 2001} {
$^{\dagger}${\tt obers@phys.uu.nl} \\ 
$^{\ddagger}${\tt pioline@physics.harvard.edu} 
\\ $^{\star}$ Talk presented by NO at 1st workshop of the RTN network 
{\it The quantum structure of spacetime and the geometric nature
of fundamental interactions}.   
Work supported in part by the Stichting FOM and
 European Research Training Network
RTN1-1999-00116.}

\clearpage

\renewcommand{\baselinestretch}{.9}\rm
\renewcommand{\baselinestretch}{1.2}\rm    

\section{Introduction}

The discovery of non-perturbative dualities and symmetries
 of string theory over the last five years,
has opened up the possibility of obtaining quantities, such as
spectrum and amplitudes, that are exact for all values of the
string coupling $g_s$. In particular, the computation of exact
amplitudes in string theory has on the one hand allowed to test
the various conjectured dualities, and on the other hand enabled
us to compute for the first time instanton corrections to the
effective action. This in turn makes it possible to address the
challenging problem of understanding the rules of semi-classical
instanton calculus in string theory.

One of the central ingredients in this development has been
supersymmetry, which protects special ``BPS'' states in the spectrum,
invariant under part of the supersymmetry charges, from quantum
corrections.
Likewise, some special ``BPS-saturated'' terms in the effective
action do not receive perturbative corrections beyond a
certain order (typically one-loop). Imposing in addition invariance
under the duality symmetry groups, often allows to determine these
amplitudes in terms of automorphic forms, which generalize the familiar
$Sl(2,Z)$ modular forms to higher arithmetic groups.

The coupling $f_{{\cal{O}}}$ to a higher dimension operator
${\cal{O}}$ in the effective action of some compactified string
theory is in general some function on the moduli space, which
includes the dilaton, along with the moduli describing the size
and shape of the compactification manifold, as well as possible
other background fields (e.g. the vevs of the RR gauge
potentials). For a BPS-saturated amplitude this coupling has
a weak-coupling expansion of the form
\begin{equation}
f_{{\cal{O}}} \sim \mbox{perturbative} + \mbox{instanton}
\end{equation}
with only a finite number of perturbative terms. From the
non-perturbative part, describing the contributions of instantons,
one may then read off the semiclassical configurations that are
present, and e.g. the corresponding instanton measure.

 For the methodology behind the computation of
exact amplitudes it is useful to distinguish between two types of
dualities. The first, and most powerful, type is duality {\it
symmetry}, where distinct regimes of the the same theory are
related. This is referred to as U-duality and includes
perturbative T-duality as well as symmetries relating strong with
weak string coupling. Examples for theories with 32
supersymmetries include the $Sl(2,\Zint)$ duality of
(uncompactified) type IIB string theory, and the more general
$\exc (\Zint)$ duality symmetry
 of type II
theory on the $d$-torus $T^d$ (or equivalently, M-theory on
$T^{d+1}$) \cite{Hull:1995ys,Townsend:1995kk,Witten:1995ex} (see
\cite{Obers:1998fb} for a review of U-duality). In the case of
toroidal compactification of the heterotic string, which has 16
supersymmetries, it is also known that the $SO(d,d+16,\Zint)$
T-duality symmetry for $d\leq 5$ is enhanced to a non-perturbative
$SO(6,22,\Zint) \times Sl(2,\Zint)$ for $d=6$ and $SO(8,24,\Zint)$
for $d=7$. Finally, type IIB on $K_3$ has an $SO(5,21,\Zint)$
U-duality symmetry. These cases are to be contrasted with
string-string dualities, which relate two different string
theories in distinct regimes. The most prominent examples, all
having 16 supersymmetries, are heterotic-type I duality
\cite{Polchinski:1996df}, which already holds in ten dimensions,
as well as the duality between heterotic on $T^4$ and type IIA on
$K_3$ \cite{Sen:1995cj}.

When the theory has a U-duality symmetry, supersymmetry
constraints in the form of second order differential equations
together with constraints coming from duality invariance and the
known leading perturbative behavior have been used to obtain exact
BPS-saturated couplings. The paradigm in theories with 32
supersymmetries has been the $R^4$ coupling, which e.g. for
ten-dimensional type IIB has been shown to be equal to a certain
(non-holomorphic) form of weight 0 of the S-duality group, namely
the Eisenstein series of order 3/2 
\cite{Green:1997tv,Pioline:1998mn}. 
A perturbative expansion of the function exhibits the known tree and
one-loop contribution, as well as an infinite series of
exponentially suppressed terms attributed to D-instantons. This
was generalized for toroidal compactifications of type II theory
to Eisenstein series of the corresponding U-duality groups
\cite{Green:1997di,Kiritsis:1997em,Pioline:1997pu,Obers:1999um}.

In the case of string-string duality one cannot resort to such
methods, but one has to depend on non-renormalization theorems
stating that certain couplings are perturbatively (typically
one-loop) exact. Under the duality these couplings then translate
into non-perturbative results on the dual side, mapping
world-sheet instantons to space-time instantons. The canonical
examples, in theories with 16 supersymmetries, are the $R^2$
coupling \cite{Harvey:1996ir,Gregori:1997hi}, which is believed to
be one-loop exact in type II on $K_3 \times T^2$, and the $F^4$
coupling
\cite{Bachas:1997mc,Kiritsis:1997hf,Antoniadis:1997zt,Lerche:1998nx,Lerche:1998gz,Kiritsis:2000zi},
believed to be one-loop exact in toroidally compactified heterotic
string up to four dimensions.

An interesting hybrid case consists of the duality in three
dimensions between heterotic on $T^7$ and type IIA on $K_3 \times
T^3$, each of which also has an $SO(8,24,\Zint)$ U-duality
symmetry \cite{Sen:1995wr}. The latter can be used to covariantize
the one-loop heterotic result into an exact expression
\cite{Obers:2000ta}, which can then be used in turn to obtain to
exact threshold on the type IIA side using string-string duality.

In this lecture, mainly based on
\cite{Kiritsis:2000zi,Obers:2000ta}, we focus on the last two
cases above, referring to \cite{Obers:1998rn} for a summary of the
use of U-duality symmetry and Eisenstein series in determining the
exact $R^4$ couplings in toroidal compactifications of type II
strings and M-theory. Section 2 discusses the moduli space of
heterotic string theory on $T^d$ and its relation via triality to
type II string theory on $K_3 \times T^{d-4}$. In Section 3 we
present a non-trivial test of heterotic-type IIA duality by
matching the one-loop $F^4$ heterotic amplitude in six dimensions
to the dual tree-level type IIA result. Then in Section 4, we
obtain from the $F^4$ amplitude in $D=4$ and 5, non-perturbative
D- and NS5-brane instanton effects in type IIA. Finally, Section 5
discusses the special three-dimensional case in which one finds
instanton effects on both of the dual sides, revealing in
particular heterotic 5-brane instantons as well as  KK5-brane
instantons on either side.

\section{Heterotic and IIA moduli spaces and triality \label{mod} }

The comparison of spectrum and amplitudes in two dual string
theories requires a proper understanding of the corresponding
moduli spaces and the precise mapping between them. Therefore, we
start by recalling some basic facts about the heterotic moduli
space for toroidal compactifications. For simplicity we consider
mainly compactifications on square tori, even though the full
duality becomes apparent only when including the Wilson lines and
vielbein moduli on the heterotic side and the Ramond backgrounds
on the type-II side.

We first consider the $E_8\times E_8$ or $SO(32)$ heterotic string
theory compactified on a torus $T^{d}$. For $d\leq 5$, the moduli
space takes the form
\begin{equation}
\label{hetmod} \Real^+ \times \so{d}{d+16} /  SO(d,d+16,\Zint) \ ,
\end{equation}
where the first factor is parameterized by the T-duality invariant
dilaton $\phi_{10-d}$ related to the ten-dimensional heterotic
coupling $g_{\rm H}$ by $e^{-2\phi_{10-d}}=V_d / (g_{\rm H}^2
l_{\rm H}^d)$, with $V_d$ the volume of the $d$-torus; the second
factor is the standard Narain moduli space, describing the metric
$g$ and B-field $b$ on the torus $T^d$, together with the Wilson
lines $y$ of the 16 U(1) gauge fields (in the Cartan torus of the
ten-dimensional gauge group) along the $d$ circles of the torus.
The right action of the discrete group $SO(d,d+16,\Zint)$ reflects
the invariance under T-duality. This moduli space can be
parameterized by a viel-bein $e_{\rm H}\in SO(d,d+16,\Real)$ in
the upper-triangular (Iwasawa) gauge. In particular, the dilatonic
moduli parameterizing the Abelian (diagonal) part of the  vielbein
are the radii of compactification
\begin{equation}
x_{i=1\ldots d} =( R_1/\lh, \dots, R_d/\lh ) \ .
\end{equation}

In order to determine the mapping of moduli to the dual
descriptions, we will compare the BPS mass formula on both sides.
For perturbative heterotic BPS states, it is simply given by
\begin{equation}
\label{masshet} {\cal M} ^2 = \frac{1}{\lh ^2} Q^t
(M_{d,d+16}-\eta) Q\ ,
\end{equation}
where $Q=(m^i,q^I,n_i)$ is the vector of momenta, charges and
windings and $M_{d,d+16}=e^t_{\rm H} e_{\rm H}$ in terms of the
viel-bein. The charges $q^I$,$I=1\dots 16$ take values in the even
self dual lattice $E_8\oplus E_8$ or $D_{16}$.

This description of the moduli space is quite complete for
compactification down to 5 dimensions. For $D=4$, however, the
$\Real^+$ factor in \eqref{hetmod} is enhanced to $U(1)\backslash
Sl(2,\Real)$, parameterized by a complex parameter $S=a+i/\ghf^2$,
where $1/\ghf^2=V_6/(\gh^2\lh^6)$ is the four-dimensional string
coupling constant and $a$ the scalar dual to the Neveu-Schwarz
two-form in four dimensions. The action of $Sl(2,\Zint)$ on $S$ is
conjectured to be a non-perturbative symmetry \cite{Font:1990gx}
of the heterotic string compactified on $T^6$. Further enhancement
in $D=3$ is due to the dualization  of the 30 $U(1)$ gauge fields
into scalars, which are then unified together with the dilaton and
$SO(7,23)$ scalars into a $\so{8}{24}$ symmetric manifold, acted
upon by the U-duality group $SO(8,24,\Zint)$ \cite{Sen:1995wr}. To
identify this subgroup of $SO(8,24)$ which remains as a quantum
symmetry, it is useful to translate the action of the Weyl
generator $\ghf\to 1/\ghf, \lh\to \ghf^2\lh$ of the $Sl(2,\Zint)$
S-duality in $D=4$ in terms of the three-dimensional variables
using $1/\ght^2=V_7/(\gh^2 \lh^7)$. The result is an exchange of
$1/\ght^2$ and $R_7/\lh$, so that using the Weyl group of
$Sl(7)\subset SO(7,23)$, we can transform $1/\ght^2$ into
$R_i/\lh$ for any radius of $T^7$. This implies that we can think
of the $SO(8,24)$ scalars as the moduli of an heterotic
compactification on $T^7\times S^1$, where the radius of $S_1$ is
given by
\begin{equation}
\label{r8def} R_8/\lh  = 1/\ght^2 \ .
\end{equation}
This compact circle therefore appears as a dynamically generated
dimension decompactifying at weak coupling. Note that this is {\it
not} the usual M-theory direction, whose radius is instead
$R_{11}=\gh \lh$. The dilatonic moduli parameterizing the Abelian
part of $SO(8,24)$ are therefore
\begin{equation}
\label{hetdil} x_{i=1\dots 8}= ( R_1/\lh, \dots, R_7/\lh,
1/\ght^2) \ .
\end{equation}

We now turn to the dual compactifications, for which we only
present  the salient features, referring to \cite{Kiritsis:2000zi}
for a detailed treatment of the moduli spaces and their mapping to
the heterotic theory. We also constrain ourselves to the case $d
\geq 4$, in which case the heterotic theory on $T^d$ is dual to
type IIA on
 $K_3 \times T^{d-4}$, and further restrict to the $T^4/\Zint_2$
 orbifold point of $K_3$.
 Starting with $d=4$, the duality map
 is given by \cite{Sen:1995cj}
\begin{equation}
\lh=\giis \lii\ \ ,\quad \giis=\frac{1}{\ghs}\ ,\quad \left(
\frac{R_1}{l_{\rm H}} \right)^2 = \frac{V_{K_3}}{\lii ^4}\ ,
\label{hetiia}
\end{equation}
which can be obtained by identifying the IIA NS5-brane on $K_3$
with the fundamental heterotic string, and the type IIA D0-brane
with a heterotic Kaluza-Klein state along the circle of radius
$R_1$ in $T^4$.

The relation between the heterotic and type II moduli can be
obtained by studying the BPS spectrum. On the heterotic side, the
BPS states are Kaluza-Klein and winding states transforming as a
{\it vector} of $SO(4,4,\Zint)$, and possibly charged under the 16
$U(1)$ gauge fields. On the type IIA side, a set of BPS states is
certainly given by the D0-, D2- and D4-branes wrapped on the even
cycles of $T^4$, which are invariant under the $\Zint_2$
involution. These states transform as a {\it conjugate spinor} of
the T-duality group $SO(4,4,\Zint)$, as D-branes should
\cite{Obers:1998fb}. We thus find that the heterotic $g/\lh^2,b$
and type IIA $G/\lii^2,B$ moduli should be related by $SO(4,4)$
{\it triality} \cite{Nahm:1999ps,Kiritsis:2000zi}, which exchanges
the vector and conjugate spinor representations. In particular, at
the $T^4/\Zint_2$ orbifold point with a square $T^4$, this reduces
to
\begin{equation}
\label{trial} R_1\vert_{\rm H}={\sqrt{R_1 R_2 R_3
R_4}}_{\vert\iia}\sp R_i\vert_{\rm H}={\sqrt\frac{R_1 R_i}{ R_j
R_k}}_{\vert\iia}\sp i,j,k=2,3,4 \ , 
\end{equation}
where the radii are measured in the respective string length units. 
In logarithmic units, we recognize the standard triality matrix acting in
the $SO(8)$ weight space.

 There are also
D2-brane states wrapped on the collapsed spheres at the sixteen
orbifold singularities \cite{Douglas:1996sw}, and charged under
the corresponding $U(1)$ fields. These are to be identified with
the charged BPS states on the heterotic side, and their masses are
matched by choosing the Wilson lines as \cite{Bergman:1999kq}
\begin{equation}
\label{hety} y=\frac{1}{2} \left( \begin{array}{cccc} 0101 &0101
&0101 &0101\\ 0000 &0000 &1111 &1111\\ 0000 &1111 &0000 &1111\\
0011 &0011 &0011 &0011
\end{array} \right) \ .
\end{equation}
This can also be derived by realizing that the Wilson lines along
the first circle in $T^4$ map to the B-field fluxes on the
collapsed two-spheres, which have been shown to be half a unit in
order for the conformal field theory to be non-singular
\cite{Aspinwall:1996mn}. If we instead put this Wilson line to
zero, we recover a gauge symmetry $SO(4)^8=SU(2)^{16}$, as
appropriate for the 16 $A_1$ singularities of $T^4/\Zint_2$. This
choice is relevant for M-theory compactified on $K_3$ at the
$\Zint_2$ orbifold point. If we further omit the Wilson lines on
the 2nd (resp 2nd and 3rd) circles, the gauge symmetry is enlarged
to $SO(8)^4$ (resp. $SO(16)^2$), which are relevant for F-theory
on $K_3$ and type I' respectively. These relations explain why our
results can easily been applied to these higher dimensional cases
as well \cite{Kiritsis:2000zi}.

For the cases $d > 4$, we note that the $T^{d-4}$ torus in the
compactification of type IIA on $K_3 \times T^{d-4}$ is inert
under heterotic-type II duality so that we can use the 6D duality
map \eqref{hetiia} to obtain the relation between this part of the
compactification manifold on the two dual sides. For $d=5$, this
gives that the extra circles of compactification on the two dual
sides are related by
\begin{equation}
\label{5Dmap} R_{\rm H} = R_{\rm A}/\giis \ .
\end{equation}
 When $d=6$, we obtain the mapping
\begin{equation}
\label{st} S_{\rm H}=T_{\iia}\ , \quad T_{\rm H}=S_{\iia}\ ,\quad
U_{\rm H}=U_{\iia}
\end{equation}
between the four-dimensional couplings $S=a+i/g_4^2$, K{\"a}hler
class $T$ and complex structure $U$ of the extra%
\footnote{By
further T-duality in one direction one may go to type IIB. For
$d=5$ this gives the SO(5,21) moduli of IIB/$K_3$ by
decompactifying, while for $d=6$ this induces $U \leftrightarrow
T$ interchange in \eqref{st}.} $T^2$. Finally, for the $d=7$ case,
i.e. type IIA on $K_3 \times T^3$, the set of dilatonic moduli
\eqref{hetdil} of the $SO(8,24)$ coset representative is mapped to
\begin{equation}
\label{iidil} x_{i=1\dots 8}= \left( \sqrt{R_1 R_2 R_3 R_4},
\sqrt\frac{R_1 R_2}{ R_3 R_4}, \sqrt\frac{R_1 R_3}{ R_2 R_4},
\sqrt\frac{R_1 R_4}{ R_2 R_3},
\frac{R_5}{\giis},\frac{R_6}{\giis},\frac{R_7}{\giis}, \frac{R_5
R_6 R_7}{\giis} \right) \ .
\end{equation}
These moduli are not the standard $SO(4,4)\times SO(3,3)$ that
arise from the reduction of the six-dimensional type II theory on
$T^4/\Zint_2 \times T^3$. The latter can however be reached by
performing, in analogy with \eqref{trial}, an $SO(4,4)$ triality
acting on both $T^4/\Zint_2$ moduli, and another $SO(4,4)$
triality on the non-perturbative $T^4$ torus \cite{Obers:2000ta}.
This gives the more familiar parameterization
\begin{equation}
\label{iidi2} y_{i=1\dots 8}= \left( R_1 , \dots, R_7, 1/\giit^2 
 \right) \ , 
\end{equation}
so that type IIA theory compactified on $K_3\times T^3$ also
appears to have a dynamically generated dimension, of size $\tilde R_8
=\lii/\giit^2$.

\section{Heterotic one-loop $F^4$ coupling and tree level IIA \label{tree}
}

Our first check of duality will be the matching of the $F^4$
couplings in $D=6$ on the heterotic and type II side. On both
sides there are 24 gauge fields. On the heterotic side, there are
(0,16) vectors originating from the Cartan torus of the
ten-dimensional gauge group and (4,0)+(0,4) from the reduction of the
metric and antisymmetric tensor on $T^4$, corresponding to 4
vectors and 4 graviphotons. On the type IIA side, these gauge
fields originate from the Ramond-Ramond sector. The 4 graviphotons
can be understood as the reduction of the 4-form field strength in
$D=10$ on the 3 self-dual cycles of $K_3$ together with the
ten-dimensional 2-form field-strength, whereas the 20 vector
multiplets come from the 4-form in $D=10$ on the 19 anti-self-dual
cycles together with the 6-form field strength $K_3$ itself. At
the $T^4/\Zint_2$ orbifold point, the gauge fields split into
(4,4) from the untwisted sectors and (0,16) (at each of the fixed
points) from the twisted sectors.

For simplicity we will focus here on the four-point amplitudes of
the  (0,16) vectors, referring to \cite{Kiritsis:2000zi} for
details on the other gauge fields.
 There are several
arguments that support the fact that these $F^4$ couplings are
non-renormalized
\cite{Yasuda:1988bu,Tseytlin:1996cg,Bachas:1996bp,Bachas:1997mc}
in the heterotic string beyond one-loop for $D \geq 4$. Firstly,
in ten dimensions these terms are related by supersymmetry to
CP-odd couplings such as $B \wedge {\rm Tr} F^4$, which should
receive no corrections beyond one-loop for anomaly cancellation. A
more explicit proof can be given at the level of string amplitudes
\cite{Yasuda:1988bu}. Secondly, the only heterotic half-BPS
instanton is the heterotic 5-brane, which needs a six-cycle to
wrap in order to give a finite action instanton effect. So for
$d<6$ there can be no non-perturbative contributions beyond the
one-loop result. Third, it is consistent with the factorization of
the moduli space \eqref{hetmod} and the T-duality symmetry
$O(d,d+16,\Zint)$ to assume that $t_8 \Tr F^4$ couplings are given
at one-loop only and hence independent of the $\Real^+$ factor. In
$d=6$ it is plausible that supersymmetry prevents the mixing of
the $Sl(2,\Real)$ dilaton factor with the Narain moduli in $F^4$
couplings, in the same way as neutral hypermultiplets decouple
from vector multiplets in $N=2$ supergravity, and prevents
corrections from NS5-brane instantons \cite{Kiritsis:1999ss}. For
$d=7$, U-duality mixes the dilaton with the Narain moduli, so that
a similar statement cannot hold. Gauge fields being Poincar\'e
dual to scalars in 3 dimensions, the $F^4$ couplings translate
into four-derivative scalar couplings, and should receive
non-perturbative corrections, as we will see explicitly in Section
\ref{3D}.

Since the $F^4$ amplitude in toroidally compactified heterotic
string theory is half-BPS saturated, the left-moving part of the
four-gauge-boson (or graviton) amplitude
 merely provides the kinematic structure, whereas
the right-moving currents reduce to their zero-mode part. The
scattering amplitude between four vector multiplets at the
one-loop level is given by the ``elliptic genus''
\cite{Lerche:1987sg}
\begin{equation}
\label{f4h} \Delta^{\rm Het,1-loop}_{F^4}= \lh^{d-2} \int_{\cal F}
\frac{d^2\tau}{\tau_2^2} \frac{p_R^4\cdot
Z_{d,d+16}(g/\lh^2,b,y)}{\bar\eta^{24}} \ ,
\end{equation}
where we have stated the result for general $d$.
 Here $\lh$ is the
heterotic string length, and is reinstated on dimensional grounds.
$Z_{d,d+16}$ denotes the partition function of the heterotic even
self-dual lattice of signature $(d,d+16)$, parameterized by the
metric $g$, Kalb-Ramond field $b$ and Wilson lines $y$. $p_R$ has
modular weight (0,1) and inserts the right-moving momentum
corresponding to the choice of gauge boson $F$, and
$1/\eta^{24}=1/q+24+\dots$ is the contribution of the 24
right-moving oscillators that generate the Hagedorn density of
half-BPS states in the perturbative spectrum of the heterotic
string. We remark here that amplitudes involving gravitons are
similar with an insertion of an (almost) holomorphic form of
weight (4,0), $\alpha E_4 + \beta \hat E_2^2$ instead of $p_R^4$.
Amplitudes involving graviphotons have more powers of lattice
momenta.

Considering now the case $d=4$, we find using the duality
relations \eqref{hetiia} with type IIA on $K_3$ and taking into account
the particular normalization of the type II Ramond fields, that
 \eqref{f4h} translates into a
tree-level type IIA result. On the other hand, it is still given
by a modular integral on the fundamental domain of the upper-half
plane, which is usually characteristic of one-loop amplitudes. The
resolution of this paradox is that on the type IIA side, the gauge
fields dual to the $(0,16)$ heterotic ones originate from the
twisted sectors of the orbifold. The correlator of four $\Zint_2$
twist fields on the sphere can be re-expressed as the correlator
of single-valued fields on the double cover of the sphere, which
is a torus \cite{Dixon:1987qv}; its modulus depends on the
relative position of the four vertices, and hence should be
integrated over, in qualitative agreement with the heterotic side.
A careful computation yields the tree-level type IIA result
\begin{equation}
\label{a4iig} \Delta^{\iia,{\rm tree}}_{F^4}= \frac{1}{\gii^2} \gii^4
\frac{\lii^6}{V_{K_3}}
 \int_{\cal F} \frac{d^2\tau}{\tau_2^2} Z_{4,4}(G/\lii^2,B) \ ,
\end{equation}
where the factors of $\gii$ correspond to the tree-level weight
and the normalization of the Ramond fields respectively and $G$
and $B$ are the moduli of the $T^4$ on the IIA side. Here we have
focused for simplicity on a particular choice of $(0,16)$ fields:
in general, \eqref{a4iig} involves a shifted lattice sum
integrated on a six-fold cover ${\cal F}_2$ of the fundamental
domain ${\cal F}$.
 Note in particular, that the oscillators have dropped, in agreement with
 the fact that this amplitude
should be half-BPS saturated. The normalization factor
$\lii^6/V_{K_3}$  has been chosen so as to agree with the
heterotic result.

Still the type IIA result \eqref{a4iig} is not quite of the same form
as \eqref{f4h}. For one thing, the type IIA result, being a
half-BPS saturated coupling, does not involve any oscillators, in
contrast to the heterotic side. For another, the $\so{4}{4}$
moduli $G/\lii^2,B$ are not the same as the heterotic $g/\lh^2,b$.
In order to reconcile the two, we need two further ingredients.

\noindent {\bf Hecke identities}.
Since we are interested in comparing the dual theories at the
orbifold point of $K_3$, we need to implement the choice of Wilson
lines \eqref{hety} on the heterotic side. These are best described
 \cite{Gava:1999ky} in terms of a $(\Zint_2)^4$ freely acting orbifold, so that
\begin{equation}
\label{z420het} Z_{4,20} = \frac{1}{2^4} \sum_{h,g}
Z_{4,4}\ar{h}{g} \bar\Theta_{16}\ar{h}{g}  \ ,
\end{equation}
where $g$ and $h$ run from 0 to 15 and are best seen as four-digit
binary numbers; $h$ labels the twisted sector while the summation
over $g$ implements the orbifold projection in that sector. The
blocks $Z_{4,4}\ar{h}{g}$ are partition functions of (4,4)
lattices with half-integer shifts, and $\bar\Theta_{16}\ar{h}{g}$
are antiholomorphic conformal characters.
 The operator $\bar Q^4 = p_R^4 $
only acts on the latter. Using techniques first developed in
\cite{Mayr:1993mq}, it can be shown
\cite{Lerche:1998nx,Kiritsis:2000zi} that the conformal blocks
$\Phi\ar{h}{g}=Q^4 \Theta_{16}\ar{h}{g}/\eta^{24}$ occurring in
the modular integral can be replaced by two-thirds their image
$\lambda$ under the Hecke operator
\begin{eqnarray}
H_{\Gamma_2^-}.\Phi(\tau)=\frac{1}{2} \left(
 \Phi\left(-\frac{1}{2\tau}\right) +
\Phi\left(\frac{\tau}{2}\right)+ \Phi\left(\frac{\tau+1}{2}\right)
\right)
\end{eqnarray}
provided this image is a constant real number:
\begin{equation}
\label{hecke} H_{\Gamma_2^-}\cdot \left[ \frac{Q^4
\Theta_{16}\ar{0}{1}}{\eta^{24}} \right]=\lambda \ .
\end{equation}
The relation \eqref{hecke} indeed holds for all the conformal
blocks of interest in this construction. The modular integral thus
reduces to
\begin{equation}
\label{f4hr} A^{\rm Het}_{F^4}=   \lh^2
\int_{\cal F} \frac{d^2\tau}{\tau_2^2}
 Z_{4,4}(g/\lh^2,b)
\end{equation}
and the Hagedorn density of half-BPS states in \eqref{f4h} has
thus cancelled.

\noindent{\bf Triality.} The last step needed to identify the type
IIA and heterotic result is to understand how the triality
transformation \eqref{trial} relating the moduli on the dual
sides, equates the integrals of the partition function
$Z_{4,4}(g/\lh^2,b)$ and $Z_{4,4}(G/\lii^2,B)$ on the fundamental
domain of the upper-half plane. It is easy to convince oneself
that such an equality cannot hold at the level of integrands, by
looking at some decompactification limits. However, it has been
shown that such modular integrals could be represented either as
Eisenstein series for the T-duality group $SO(4,4,\Zint)$, in the
{\it vector} or (conjugate) {\it spinor} representations, thanks
to the identity \cite{Obers:1999um}:
\begin{equation}
\pi \int_{\cal F} \frac{d^2\tau}{\tau_2^2}Z_{4,4}
=\eis{SO(4,4,\Zint)}{V}{s=1}=\eis{SO(4,4,\Zint)}{S}{s=1}
=\eis{SO(4,4,\Zint)}{C}{s=1} \ .
\end{equation}
This implies the invariance of the modular integral of
$Z_{4,4}(g/l_s^2,b)$ under triality transformation of the moduli,
which completes the argument \cite{Kiritsis:2000zi}  that in six dimensions the one-loop
$F^4$ heterotic coupling equals the tree-level $F^4$ coupling in
IIA.

In the above, we have focused mainly on the case of four identical
gauge fields in (0,16). Similarly, one can consider two groups of
two identical gauge fields or four distinct ones. Also in these
cases can one make a successful comparison between the heterotic
and type IIA side, though a precise match requires the exact
identification of the gauge fields on either side, which is still
an open problem. It would be also interesting to understand how
the duality holds at other orbifold points of $K_3$, since naively
the correlator of $\Zint_n$ twist fields on the sphere involves
higher genus Riemann surfaces, albeit of a very symmetric type.

\section{D-instantons and NS5-brane instantons in IIA \label{45D} }

Having reproduced the type IIA tree-level $F^4$ coupling in 6
dimensions from the heterotic one-loop amplitude, we now would
like to use the duality map in lower dimensions to show some
non-trivial consequences of string-string duality. The absence of
non-perturbative corrections to $F^4$ on the type IIA side in
$D=6$ is easily understood as follows: Such instanton effects
should arise from half-BPS solitonic solutions, with their entire
(Euclidean) world-volume wrapped on supersymmetric cycles of the
same dimensionality. Since, type IIA has only even D-branes, these
should wrap odd-dimensional supersymmetric cycles, which are not
present in $K_3$, which has only 0,2 and 4-cycles. Moreover,
NS5-brane (KK5-brane) instantons would require six- (seven-) cycles,
and hence cannot occur until $D=4$ ($D=3$). On the other hand,
this implies that for compactification on $K_3 \times T^{d-4}$,
with $d > 4$, we would expect instanton contributions on these
general grounds. In particular, we expect for increasing $d$
the following instantons to appear:
\newline
$\bullet $ $d=5$: D0,D2,D4-brane, wrapping 0,2,4-cycle in $K_3$ and $S^1$ \newline
$\bullet $ $d=6$: NS5-brane, wrapping 4-cycle in $K_3$ and $T^2$
 \newline
 $\bullet $ $d=7$: D6,KK5-brane, wrapping 4-cycle in $K_3$ and $T^3$
 \newline In this section, we explicitly derive these effects from
 the one-loop heterotic $F^4$ coupling for the first two
 cases%
 \footnote{We note here that, by the same reasoning, in the type
IIB picture instanton configurations already exist in $D=6$,
since D-1,D1 and D3 branes can wrap the 0,2 and 4-cycles, as first
computed in \cite{Antoniadis:1997zt}. See also
\cite{Kiritsis:2000zi} for the $SO(5,21,\Zint)$ U-duality
invariant $t_{12} H^4$ coupling in that case.} above, leaving the
last case to Section \ref{3D}

The general approach to quantitatively understand these instanton
configurations from the 1-loop heterotic result at the orbifold point 
\begin{equation}
\label{del5} \Delta_{F^4}^{\rm Het} = \lh^{d-2} \int_{\F}
\frac{d^2\tau}{\tau_2^2} Z_{d,d}(g/\lh^2,b) \ ,
\end{equation}
is as follows. First, using the duality map we find
that weak coupling expansion on the IIA side corresponds to
a large volume expansion of $T^{d-4}$ on the heterotic side.
At the level of the toroidal lattice sum $Z_{d,d}$, this
large volume expansion
is best exhibited by adopting a Hamiltonian representation
for
the $Z_{4,4}$ part and a Lagrangean representation for the remaining 
$Z_{d-4,d-4}$ part. Assuming for simplicity zero Wilson lines of the
6D-gauge fields around the $T^{d-4}$,
we thus decompose
\begin{equation}
\label{zdd} Z_{d,d} =Z_{4,4} Z_{d-4,d-4} \sp Z_{4,4} =
\sum_{m_i,n^i} q^{\frac{p_L^2}{2}} \bar q^{\frac{p_R^2}{2}} \sp q
= e^{2 \pi i \tau}  \ ,
\end{equation}
\begin{equation}
\label{zpdd}
Z_{d-4,d-4} = V_{d-4} \sum_{p^\alpha, q^\alpha} e^{ -\frac{\pi}{\tau_2}
 (p^\alpha + \tau q^\alpha )
 (g_{\alpha \beta}+b_{\alpha \beta}) (p^\beta + \bar
\tau q^\beta ) } \ .
\end{equation}
Here, $m_i,n^i$ denote the momenta and windings on $T^4$ and $ 2
p_{L,R}^2 = m^t (M_{4,4}  \pm  \eta ) m$ with $m=(m_i,n^i)$, and
$M_{4,4}$ the $SO(4,4)$ moduli matrix. Then we use the method of
orbits on the last factor to evaluate the modular integral as a
sum over orbits, where each term in the orbit decomposition will
be a distinct term in the large $T^{d-4}$
 volume expansion. In particular, this shows a (finite) power
 series in the volume, corresponding to
 perturbative contributions on the dual side. The interesting part
 consists of the terms that are exponentially suppressed and are due to
 world-sheet instantons on the heterotic side, which map to
 space-time instantons on the type IIA side.

Let us see first how this applies to $D=5$, in which case it
follows from the duality relation \eqref{5Dmap} that the weak coupling
regime on the type IIA side corresponds to the large $R_{\rm H}$
expansion on the heterotic side. According to \eqref{zdd}, \eqref{zpdd} the
5D-coupling can thus be written as
\begin{equation}
\label{del5lh} \Delta_{5D}=\lh^{2} \rh \int_{\cal F}
\frac{d^{2}\tau}{\tau_2^2} \sum_{p,q} \sum_{m_i,n^i} \exp\left( -
\pi \frac {\rh^2|p-\tau q|^2 }{\lh^2 \tau_2} \right) \tau_2^2
q^{\frac{p_L^2}{2}} \bar q^{\frac{p_R^2}{2}} \ .
\end{equation}
We now apply the standard orbit decomposition method on the integers $(p,q)$,
trading the sum over $Sl(2,\Zint)$ images of $(p,q)$ for
a sum over images of the fundamental domain ${\cal F}$ 
\cite{McClain:1987id}
(see \cite{Kiritsis:1997hf,Obers:1999um} for relevant formulae).
The zero orbit gives back the six-dimensional result
\eqref{a4iig}
up to a volume factor, and reproduces the
tree-level type II contribution in 5 dimensions:
\begin{equation}
\label{del5z} \Delta_{5D}^{\rm zero}=\lh^{2} \rh \int_{\cal F}
\frac{d^{2}\tau}{\tau_2^2} Z_{4,4}(g/\lh^2,b) = R_{\rm A}
\Delta_{{\rm IIA},4D}^{\rm tree} \ .
\end{equation}
The degenerate orbit on the other hand, with representatives
$(p,0)$, can be unfolded onto the strip $|\tau_1|<1/2$. The $\tau_1$
integral then imposes the level matching condition $p_L^2-p_R^2=2m_i n^i=0$,
and the $\tau_2$ integral can be carried out in terms of Bessel
functions. In type IIA variables the result is
\begin{equation}
\label{del5d2} \Delta_{5D}^{\rm deg}=2 \giis \lii R_{\rm A}^2
\sum_{p\neq 0} \sum_{(m_i,n^i)\neq 0}
 \delta(m_i n^i)
\cdot  \frac{|p|}{\sqrt{m^t M_{4,4} m}} K_1\left( 2\pi \frac{R_{\rm
A}}{\giis\lii}
 |p| \sqrt{m^t M_{4,4} m} \right) \ ,
\end{equation}
where $M_{4,4}$ is now the mass matrix of D-brane states wrapped
on the untwisted cycles of $T^4/\Zint_2$. Given the asymptotic
behaviour $K_1(x)\sim \sqrt{\frac{\pi}{2x}}e^{-x}$, we see that
this is a sum of order $e^{-1/g_s}$ non-perturbative effects
corresponding to $N=p r$ Euclidean (anti) D-branes wrapped on
$S_1$ times a cycle of homology charges $(m_i,n^i)/r$ on $T^4$,
where $r$ is the greatest common divisor of $(m_i,n^i)$%
\footnote{For general choices of twisted gauge fields, the actual
answers involve a torus partition function $Z_{4,4} \ar{h}{g}  $
with half integer shifts. This corresponds to open Euclidean
D-branes with half integer wrappings number, as required in order
to interpolate between different fixed points.}.

It is worth pointing out a number of peculiarities of the
result \eqref{del5d2}.
First, due to the absence of a holomorphic insertion in
\eqref{del5}, all instanton effects are due to {\it untwisted}
D-branes wrapped along even cycles of $K_3$, even though we are
discussing $F^4$ couplings between fields located on the fixed
points of the orbifold. This is in contrast to the result in
four-derivative scalar couplings \cite{Antoniadis:1997zt}, where a
contribution from the whole Hagedorn density of BPS states was found.
This is an important simplification due to our choice of the
orbifold point in the $K_3$ moduli space. Second, the integration
measure corresponding to a given number of D-branes $N$ is easily
seen to be $\sum_{r|N} (1/r^2)$, where $r$ runs over the divisors
of $N$, just as in the case of D-instanton effects in theories
with 32 supersymmetries \cite{Green:1997tv,Green:1998tn}. This is
a somewhat unexpected result, since the bulk contribution to the
index for the quantum mechanics
with 8 unbroken symmetries is $1/N^2$ instead \cite{Moore:1998et},
which did arise in four-derivative scalar couplings at the enhanced symmetry
point \cite{Antoniadis:1997zt,Green:1998yf}.

Turning to the next case, $D=4$, we see from the duality map
\eqref{st} that weak coupling in type IIA corresponds to
the limit where the heterotic $T^2$ decompactifies. The study of
this decompactification limit proceeds as outlined in the general
strategy above. Performing an orbit decomposition on the integers
running in the Lagrangean representation of the $T^2$ lattice, the
zero orbit and degenerate orbit reproduce the tree-level and
D-instanton contributions on the type II side. The novelty in this
case is that there is a third orbit, namely the non-degenerate
orbit, which contributes as well. The integral on $\tau_1$ is
Gaussian, and the subsequent integral along $\tau_2$ is again
given by a Bessel function. Before carrying out this integration
exactly, it is more enlightening to determine the saddle point,
which controls the instanton effects at leading order. Using
\eqref{zdd}, \eqref{zpdd} with $d=6$, the saddle point equations are easily
found to be
\begin{equation}
\label{sad1} q^\alpha g_{\alpha \beta} (p^\beta-\tau_1 q^\beta)+ i
\tau_2 m_i n^i=0 \ ,
\end{equation}
\begin{equation}
\label{sad2} -(p^\alpha-\tau_1 q^\alpha)g_{\alpha
\beta}(p^\beta-\tau_1 q^\beta)+\tau_2^2(q^\alpha g_{\alpha \beta}
q^\beta + m^t M_{4,4}m)=0 \ ,
\end{equation}
where $p^\alpha$ and $q^\alpha$ are the integers running in the
$T^2$ lattice partition function, and should be summed over
$Sl(2,\Zint)$ orbits such that $p^1 q^2-p^2 q^1\neq 0$ only.
$g_{\alpha \beta}$ is the metric on $T^2$ in heterotic units.
These equations are easily solved and after translation to type
IIA variables correspond to a classical action
\begin{equation}
\label{4Dact}
S_{\rm cl}=2\pi \sqrt{ \frac{p^2 q^2- (pq)^2}{(q^2)^2}
\left( \frac{(q^2)^2}{\giis^4 \lii^4}
 + \frac{q^2 m^t M_{4,4} m }{\giis^2 \lii^2} + (m_i n^i)^2 \right) }
+  2\pi i \left( \frac{ (pq) (m_i n^i)}{q^2} +  pBq \right) \ . 
\end{equation}
In particular, the real part of the classical action
scales as $1/\giis^2$. The corresponding non-perturbative
effects should therefore be interpreted as coming from
$N=|p^1 q^2-p^2 q^1|$ NS5-branes wrapped on $K_3\times T^2$,
and bound to D-brane states wrapped on an even cycle of $K_3$
times a circle on $T^2$ determined by the integers $q^1,q^2$.

The explicit result of the $\tau$ integration gives
\begin{equation}
\label{4Dnd} \Delta_{4D}^{\rm n.d.}=4 \lh^{4}
\sum_{p^\alpha,q^\alpha} \sum_{m_i,n^i} \left(\frac{(q^2)^2+q^2
m^t M_{4,4} m+(m_in^i)^2}{p^2q^2-(pq)^2}\right)^{3/4} K_{3/2}
\left( \Re S_{\rm cl} \right) e^{i \Im S_{\rm cl}} \ .
\end{equation}
In particular, we may look at the contribution of pure NS5-brane
instantons, corresponding to $m_i=n^i=0$. Choosing an orbit
representatives and using the exact expression for the Bessel
function
\begin{equation}
\label{k32} K_{3/2}(x)=\sqrt{\pi/2x}\left(1+1/x \right) e^{-x}
\end{equation}
we obtain
\begin{equation}
\label{ns5cont} 
\Delta_{4D}^{\rm NS5}= 2 (\giis \lii)^4 U_2 \sum_N \mu (N)  \left(
N+\frac{1}{2\pi S_2} \right) e^{-2\pi N S_2} \left( e^{2\pi i N
S_1}+e^{-2\pi i N S_1} \right) \ ,
\end{equation}
where we used the type II variable $S=a+iV_{K_3} V_{T^2}/(\gii^2
\lii^6)$ and extracted the instanton measure
\cite{Kiritsis:2000zi}
\begin{equation}
\label{mu5} \mu (N)=\sum_{r|N} \frac{1}{r^3} \sp (\mbox{NS5-brane
on}\;\,K_3 \times T^2) \ .
\end{equation}
This result gives a prediction for the index (or
rather the bulk contribution thereto) of the world-volume
theory of the type II NS5-brane wrapped on $K_3\times T^2$. It
is a challenging problem to try and derive this
result from first principles. It is
also remarkable that, in virtue of \eqref{k32} and in contrast to
D-instantons, the NS5-instantons contributions do not seem to
receive any perturbative subcorrections beyond one-loop.

It is interesting to compare this result to the corresponding
index of the heterotic 5-brane wrapped on $T^6$, which can be extracted
from the non-perturbative $R^2$ couplings in the heterotic string
compactified on $T^6$ \cite{Hammou:1999in,Kiritsis:1999ss}.
Those can be computed by duality from
the one-loop
exact $R^2$ couplings in type  II on $K_3\times T^2$ \cite{Harvey:1996ir},
and read
\begin{equation}
 \Delta_{R^2} =\hat{\cal E}^{Sl(2,\Zint)}_{\irrep{2};s=1}
   = -\pi \log (S_2 |
\eta(S) |^4)
 = \frac{\pi^2}{3} S_2 + 2 \pi \sqrt{S_2}
\sum_N \mu(N)e^{-2\pi N S_2} \left( e^{2\pi i N S_1}+e^{-2\pi i N
S_1} \right) \ .
\end{equation}
The summation measure turns out to be different from \eqref{mu5} and
given instead by
\begin{equation}
\mu (N)=\sum_{r|N} \frac{1}{r} \sp (\mbox{Het 5-brane on}\;\, T^6) \ .
\end{equation}
It is also worthwhile to notice that there are no subleading
corrections around the instanton  in  the heterotic 5-brane case.
This is in contrast to D-instantons, for which the saddle point
approximation to the Bessel function $K_{1}$ is not exact.
 It would be interesting to have a deeper understanding
of these non-renormalization properties, possibly  using the CFT
description of the 5-brane \cite{Callan:1991at}.

\section{Heterotic/type II duality in $D=3$ \label{3D} }

We finally present the three-dimensional case, that is the duality
between heterotic on $T^7$ and type IIA on $K_3 \times T^3$. As
shown in Section \ref{mod}, this case has the interesting feature
that there is an $SO(8,24,\Zint)$ U-duality symmetry
\cite{Sen:1995wr} on each dual side, which can be used to
conjecture a non-perturbative $F^4$ amplitude \cite{Obers:2000ta}
on both the heterotic as well as the type IIA side, while in $D>3$
this amplitude was one-loop exact. The proposed  U-duality
invariant $F^4$ amplitude reproduces the known one-loop and
tree-level answers on the heterotic and type II sides
respectively. Moreover, this amplitude will
 contain instanton corrections along with the perturbative
contributions, and hence give us information in particular about
the elusive heterotic instantons. Another property is that string
theories in three dimensions possess new instanton configurations
that were not present in $D>3$, namely the Kaluza-Klein 5-brane
instantons. These instantons are obtained by tensoring a Taub-NUT
gravitational instanton asymptotic to $\Real^3\times S^1$ with a
flat $T^6$, where $T^7=S^1\times T^6$. They are the
ten-dimensional Euclidean version \cite{Townsend:1995kk} of the
Kaluza-Klein monopoles introduced in \cite{Sorkin:1983ns}.
Moreover, $D=3$ is just one step away from $D=2$, where the
U-duality group becomes infinite-dimensional, and predicts an
infinite set of particles with exotic dependence $1/g_s^{n+3}$ on
the coupling \cite{Elitzur:1997zn,Obers:1998fb}. Similar states
already appear as particles in $D=3$ with mass $1/g_s^3$.
Instantons in $D=3$ are however free of these infrared problems,
and the study of exact amplitudes in $D=3$ may shed light on the
non-perturbative spectrum.

Our starting point is the heterotic one-loop $F^4$ amplitude in
three dimensions, given by \eqref{f4h} with $d=7$, which already
incorporates most of the symmetries, namely the T-duality
$SO(7,23,\Zint)$. First, we dualize the vectors into scalars,
which is achieved by adding a Lagrange multiplier term $\phi_a
dF^a$ to the gauge kinetic term $(\lh/\ght^2) F^a
(M_{7,23})^{-1}_{ab} F^b$ and integrating out the field strength
$F^a$, yielding
$F^a =  \frac{\ght^2}{\lh} M_{7,23}^{ab} d\phi_b$.
The resulting one-loop result is then
\begin{equation}
\label{hetloop} \Delta_{(\partial \phi)^4}^{{\rm Het},\rm 1-loop}=
\lh \ght^8 \int_{\cal F} \frac{d^{2}\tau}{\tau_2^2}
\frac{p_R^4}{\bar \eta^{24}} Z_{7,23} ~t_4 (M_{7,23}\partial
\phi)^4 \ ,
\end{equation}
where $t_4$ is the tensor $t_8$ with pairs of indices raised with the
$\e_3$ antisymmetric tensor.
This result can now be covariantized under U-duality, leading to
the proposal that  the exact $(\partial \phi)^4$ threshold in heterotic
string on $T^7$, or any of its dual formulations, is given by
\begin{equation}
\label{uduala}
I_{(\partial \phi)^4} = \lp \int d^3 x \sqrt{g} \int_{\cal F}
\frac{d^2\tau}{\tau_2^2}
\frac{Z_{8,24}(g/\lh^2,b,\phi,\ght^2)}{\bar \eta^{24}}
~t_8 [  (e_{8,24}^{-1} \partial_{\mu} e_{8,24})_{ia} ~ p_R^{a}] ^4 \ ,
\end{equation}
where $\lp = \ght^2 \lh$ is the three-dimensional (U-duality invariant)
Planck length. In this
expression,  $Z_{8,24}$ is the Theta function of the non-perturbative
$\Gamma_{8,24}$ lattice, which is invariant
 under both U-duality $SO(8,24,\Zint)$ and
$Sl(2,\Zint)$ modular transformations of $\tau$.
$e_{8,24}^{-1}\partial_\mu e_{8,24}$ is the left-invariant
one-form on the coset $[SO(8)\times SO(24)]\backslash SO(8,24)$,
pulled-back to space-time (see \cite{Obers:2000ta} for further
details). This structure is also the one that arises in a one-loop
four-scalar amplitude as shown in \cite{Antoniadis:1997zt}. The
conjecture \eqref{uduala} satisfies the following criteria:
\newline
$\bullet$ It is $SO(8,24,\Zint)$ invariant by construction;
\newline
$\bullet$  It correctly reproduces the heterotic 1-loop coupling;
\newline
$\bullet$  The non-perturbative contributions come from heterotic
5-branes and KK5-branes, which are the expected ones in $D=3$;
\newline
$\bullet $ The result decompactifies to the known purely perturbative
   result in $D \geq 4$; \newline
$\bullet$ Via heterotic/type II and heterotic/type I
duality, the corresponding amplitude in type II and I shows
the correct perturbative terms and the expected instanton
corrections.

To prove these claims, we will restrict ourselves as in Section
\ref{45D} to the particular subspace of moduli space, corresponding to
the $T^4/\Zint_2\times T^3$ orbifold point on the type II side.
As in Section \ref{tree}, for that choice of Borel moduli, the
Dedekind function $\bar \eta^{24}$
in \eqref{uduala} cancels against the action of $p_R^4$ on
the $D_{16}$ part of the lattice, and we are left with the
simpler expression
\begin{equation}
\label{np824}
\Delta_{(\partial \phi)^4} =  \lp  \int_{\cal F}
\frac{d^2\tau}{\tau_2^2} Z_{8,8}(g/\lh^2,b,\ght^2) \ ,
\end{equation}
where the partition function $Z_{8,8}$ now runs over the lattice
$\Gamma_{1,1}^8$ only.
In proving the assertions above, we will again make extensive use
of the weak-coupling expansion method
 outlined  at the beginning of Section \ref{45D}.

\noindent {\bf Heterotic instanton expansion}. We first notice
that defining $1/\ght^2 = R_8/\lh$ as in \eqref{r8def}, the weak
coupling expansion of the result \eqref{np824}  becomes a large
$R_8$ expansion. Consequently, we adopt a Lagrangean
representation for the $S^1$ part and a Hamiltonian representation
for the remainder:
\begin{equation}
\label{del7lh} \Delta_{(\partial \phi)^4}= \lp \frac{R_8}{\lh} \int_{\cal F}
\frac{d^{2}\tau}{\tau_2^2} \sum_{p,q} \sum_{v} \exp\left( -
\pi \frac {R_8^2|p-\tau q|^2 }{\lh^2 \tau_2} \right) \tau_2^{7/2}
 q^{\frac{p_L^2}{2}} \bar q^{\frac{p_R^2}{2}} \ ,
\end{equation}
where $v=(m_i,n^i)$ now denotes the $7+7$ perturbative momenta and
windings. Applying the same orbit decomposition as below
Eq.\eqref{del5lh}, we find that the zero orbit gives back the
perturbative result \eqref{hetloop}
\begin{equation}
\label{del7z} \Delta_{(\partial \phi)^4}^{\rm zero}=\lh \int_{\cal
F} \frac{d^{2}\tau}{\tau_2^2}  Z_{7,7} \ ,
\end{equation}
while the degenerate orbit gives
\begin{equation}
\label{del7d} \Delta_{(\partial \phi)^4}^{\rm deg}=2 \lh
\sum_{p\neq 0} \sum_{v \neq 0}  \delta(v^t \eta_{7,7} v ) \left(
\frac{p^2}{\ght^4 v^t M_{7,7} v} \right)^{5/4} K_{5/2}\left(
\frac{2 \pi}{\ght^2} |p| \sqrt{v^t M_{7,7} v} \right) \ .
\end{equation}
{}From the expansion of the Bessel function
\begin{equation}
\label{bes} K_{5/2}(x) = \sqrt{\pi/ 2x } [ 1 + 3/x + 3/x^2] e^{-x}
\end{equation}
we see that these are non-perturbative contributions with
classical action
\begin{equation}
 \Re (S_{\rm cl}) = \frac{2 \pi}{\ght^2} |p|
\sqrt{v^t M_{7,7} v} \ .
\end{equation}
Choosing for $v$ either
``momentum'' charges or ``winding'' charges, we find an action
\begin{equation}
\frac{1}{\ght^2} \frac{\lh}{R_i} = \frac{V_6}{\gh^2 \lh^6} \sp
\frac{1}{\ght^2}  \frac{R_i}{\lh} = \frac{V_6 R_i^2}{\gh^2 \lh^8} \ ,
\end{equation}
which identifies these instantons as heterotic 5 branes and KK5-branes
respectively, wrapped on a $T^6$ inside $T^7$. The summation measure
for these effects is easily extracted, and yields
\begin{equation}
\label{kk5m}
\mu_{\rm Het} (N) = \sum_{d|N} \frac{1}{d^5} \ ,
\end{equation}
where $N=gcd(p,m_i,n^i)$. We also note that the Bessel function $K_{5/2}$
in \eqref{bes}
exhibits only two subleading terms beyond the saddle point approximation,
so that these instanton contributions do not receive
any corrections beyond two loops.

Since NS5-brane instantons appear in the three-dimensional
$(\partial \phi)^4$ result, one may wonder how they can not
contribute in four-dimensions, where the $F^4$ threshold has been
argued to be purely one-loop
\cite{Kiritsis:1999ss,Kiritsis:2000zi}. To verify this, one
decomposes  $Z_{8,8} = Z_{2,2} Z_{6,6}$, where $Z_{2,2}$ stands
for the lattice sum on the non-perturbative two torus in the (7,8)
direction and $Z_{6,6}$ is the lattice sum for the perturbative
states in $D=4$. Decompactifying to four dimensions then corresponds to a large
volume limit of the two-torus. The orbit decomposition then shows
that only the trivial orbit contributes in the decompactification
limit, corresponding to the $D=4$ one-loop result, while both the
degenerate and non-degenerate orbits contain exponentially
suppressed terms that vanish.

\noindent {\bf Type II  instanton expansion}.
We finally consider the type II interpretation of the
 conjecture \eqref{uduala}. From the moduli identification \eqref{iidil},
 it is seen that the weak coupling
expansion on the type II side corresponds to the large
volume expansion of the non-perturbative $T^4$ in the 5--8
directions, with volume
\begin{equation}
v_4 = \frac{V_3^2}{\giis^4 \lii^6}  \ ,
\end{equation}
where $V_3 = R_5 R_6 R_7$.
Using the by now familiar method, we thus decompose
$Z_{8,8} = Z_{4,4}^{(1-4)} Z_{4,4}^{(5-8)} $ with Hamiltonian and Lagrangean
representation for the two factors respectively.
For use below, we define
the rescaled metric $G_4$ on the non-perturbative four-torus
\begin{equation}
\label{G4def}
G_4 = \giis^2 g_4 \equiv e_4^t e_4 \sp
e_4 = {\rm diag} (R_I/\lh,V_3/\lh^3) \sp I = 5,6,7 \ ,
\end{equation}
which depends on the geometric moduli only.

We can now perform again an orbit
decomposition  (see
\cite{Kiritsis:1997em,Kiritsis:1997hf} for a discussion
of the orbit decomposition for $Z_{d,d}$, $d > 2$, which
generalizes the $d=2$ case \cite{McClain:1987id}.) Then, the trivial orbit
gives
\begin{equation}
\label{delii3t} \Delta_{(\partial
\phi)^4}^{\rm zero}= \giit^2 \lii
\frac{V_3^2}{\giis^4} \int_{\cal F} \frac{d^{2}\tau}{\tau_2^2}
Z_{4,4} =
 \frac{V_3}{(\giis \lii) ^4} \Delta_{F^4}^{{\rm tree},6D} \ ,
\end{equation}
which, using \eqref{hetiia} shows the correct 3D tree-level
result, directly induced from the 6D tree-level $F^4$ coupling
\eqref{a4iig}. The degenerate orbit is evaluated as
\begin{equation}
\label{delii3d} \Delta_{(\partial \phi)^4}^{\rm deg}=2 \frac{V_3}{\giis^2
\lii^2} \sum_{p^\alpha \neq 0} \sum_{(m_i,n^i)\neq 0}  \delta(m_i n^i)
\frac{1}{\giis} \frac{\sqrt{|p^t G_4 p |} }{ \sqrt{m^t M_{4,4} m} }
K_1\left(  \frac{2\pi}{\giis} \sqrt{p^t G_4 p}  \sqrt{m^t
M_{4,4} m} \right) \ , \end{equation}
which generalizes \eqref{del5d2}. In particular, from the argument of
the Bessel function $K_1$, we recognize the contributions of
Euclidean D-branes wrapped
on an even cycle of $K_3$, times a one-cycle of $T^3$ (for
$p$ in the $5,6,7$ directions of the non-perturbative torus),
or the whole $T^3$ for $p$ in the 8th direction. The latter
case corresponds to the contributions of Euclidean D6-branes,
which start contributing in three dimensions.

For the non-degenerate orbit, the integral is dominated by
a saddle point of the same form as in \eqref{sad1}, \eqref{sad2}, with the two 
torus metric $g$ replaced by the non-perturbative four-torus metric.
Then the final integral can be expressed again in terms of the
$K_{3/2}(\Re S_{\rm cl})$ Bessel function as in \eqref{4Dnd},
where $\Re S_{\rm cl}$  is exactly of the same form as in \eqref{4Dact},
with the distinction that
all inner products of $p,q$ vectors are taken with the metric $G_4$
defined in \eqref{G4def}.
In particular, setting $m_i=n^i=0$ and switching on one charge at a time
for simplicity, we identify these non-perturbative effects as coming
from NS5-brane instantons \eqref{ns5cont} as found  already in $D=4$.
However, there are in addition genuine three-dimensional effects,
corresponding to KK5-brane instantons, with action
\begin{equation}
 \frac{V_3 R_K}{ \giis^2 \lii^4} =\frac{V_{K_3} R_I R_J R_K^2 }{\gii^2 \lii^8}
\ .
\end{equation}
 For general values of the charges $p,q,m$,
we obtain contributions from boundstates of NS5 and KK5-branes
with D-branes. It is a simple matter to derive
the summation measure for $N$ KK5-brane instantons
in type IIA/$K_3 \times T^3$, which turns out to be identical to
the one derived for NS5-brane instantons in IIA/$K_3 \times T^2$,
given in \eqref{mu5}.

\section{Conclusion and Outlook}

\begin{table}
\begin{center}
\begin{tabular}{|c||c|c|c|} \hline
% &  &  &  \\ 
     & IIA/$K_3\times T^{d-4}$ & Het/$T^d$ & I/$T^d$ \\ \hline \hline
D$p$ & 2 ($F^4$)           &       &
 \begin{tabular}{c} D1: 0 ($F^4$)\\D5: 1 ($R^2$), 5 ($F^4$) \end{tabular}
                                                    \\ \hline
NS5  & 3 ($F^4$)    & 1 ($R^2$), 5 ($F^4$) & \\ \hline KK5  & 3
($F^4$)    & 5 ($F^4$) & 5 ($F^4$) \\ \hline
\end{tabular}
\end{center}
\caption{Instanton contributions to $F^4$ and $R^2$ couplings in
theories with 16 supersymmetries. The entry denotes the exponent
$r$ appearing in the summation measure $\sum_{d|N} d^{-r}$.}
\end{table}

The $F^4$ couplings in theories with 16 supersymmetries provide a
rich setting in which to perform non-trivial tests of
string-string duality and compute new instanton effects in string
theory. We have focused in these lectures on these amplitudes in
toroidally compactified heterotic theory and its dual type IIA
compactifications in $6 \leq D \leq 3$. In fact, in the lowest
dimensional case $D=3$, one finds half-BPS instanton effects
arising from all D$p$, NS5 and KK5-branes that are present in
these theories. We emphasize that this case summarizes all other
new results as the other cases can be reached by
decompactification,
  but for pedagogical
reasons the simpler cases were discussed first.

 Many other interesting results follow by considering the
higher dimensional cases \cite{Kiritsis:2000zi}: In $D=7$, the
M-theory four-gluon amplitude for $SU(2)$ gauge bosons located at
the $A_1$ singularities of $K_3$ was obtained. In $D=8$, one
recovers the $F^4$ amplitude for $SO(8)$ gauge bosons located at
the orientifold planes of Sen's F-theory model \cite{Sen:1996vd},
also considered in \cite{Lerche:1998nx}. In $D=9$, the $F^4$
couplings at the $SO(16)\times SO(16)$ point were computed, and
shown not to involve the higher genus contact contributions found
in \cite{Bachas:1997mc,Kiritsis:1997hf}. Although we have hardly
mentioned it, the type I picture is very similar to the heterotic
one: Using heterotic/type I duality \cite{Polchinski:1996df}, the
heterotic one-loop amplitude reproduces the familiar disk and
cylinder amplitudes, together with D1-instanton effects. The
heterotic 5-brane and KK5-brane contributions turn into type I
D5-branes and KK5-branes, and their summation measure is
unaffected by the duality. The table above summarizes the
instanton summation measures for all cases known so far. It is a
challenging problem to rederive the measures for the NS5-brane and
KK5-instantons.

Among  the further open problems we mention just a few: We have
generally focused on instanton contributions coming from bound
states of a singly type of brane, but from our results it should
be possible to extract more detailed information pertaining
complicated bound states of various distinct branes. Understanding
BPS-amplitudes that break more than half of the supersymmetries is
expected to provide further insights into instanton effects, as
well as interesting generalizations of the automorphic forms that
govern half-BPS amplitudes. Another important direction is half-BPS
amplitudes in theories with yet lower supersymmetry, and computing
instanton generated superpotentials in N=1 theories. Finally, an
interesting challenge should be to consider further
compactification to two-dimensional supergravity, in which case
the U-duality symmetry is enhanced to an affine symmetry.

%\newpage
 \vskip0.5cm \noindent {\large \bf
Acknowledgements}

\smallskip
\noindent We would like to thank Elias Kiritsis for collaboration and
useful discussions on the work reviewed here. NO thanks the
organizers of the RTN workshop in Berlin for a very fruitful and
pleasant meeting.
 The
work of NO is supported in part by the Stichting FOM and the
European Commission RTN programme HPRN-CT-2000-00131.

%\begingroup\raggedright
%\newcommand{\href}[2]{#2}%{\begingroup\raggedright}
%\bibliographystyle{h-physrev3}
%\bibliographystyle{utphys}
%\bibliographystyle{JHEP}
%\bibliography{biblioniels}
%\bibliographystyle{../INPUT/utphys}
% \bibliography{../BIB/biblioniels}

\providecommand{\href}[2]{#2}\begingroup\raggedright\endgroup

\end{document}